\documentclass[a4paper,fleqn,usenatbib]{mnras}

\usepackage{newtxtext,newtxmath}

\usepackage[T1]{fontenc}
\usepackage{ae,aecompl}

\usepackage{graphicx} 
\usepackage{amsmath}  
\usepackage{amssymb}  

\usepackage{color}

\newcommand\SPIE{{Proc. SPIE}}%
\usepackage{soul}

\newcommand{\chandra}{{\it{Chandra}} \,}
\newcommand{\xmm}{{\it{XMM-Newton}} \,}
\newcommand{\nustar}{{\it{NuSTAR}} \,}
\newcommand{\wise}{{\it{WISE}} \,}
\newcommand{\ciao}{{\textsc{CIAO}} \,}
\newcommand{\heasoft}{{\textsc{HEAsoft}} \,}

\newcommand{\nustardas}{{\textsc{NuSTARDAS}} \,}

\newcommand{\heii}[1]{{\ensuremath{\mathrm{He}}}\,\textsc{ii}\:#1}
\newcommand{\oiii}[1]{{\ensuremath{\mathrm{O}}}\,\textsc{iii}}
\newcommand{\nii}[1]{{\ensuremath{\mathrm{N}}}\,\textsc{ii}}
\newcommand{\Ka}[1]{{\ensuremath{\mathrm{Fe}}} {\ensuremath{\mathrm{K}}}$\alpha$ \,}

\title[NuSTAR and Chandra observations of IC\,2497]{Joint {\it{NuSTAR}} and {\it{Chandra}} analysis of the obscured quasar in IC\,2497 - Hanny's Voorwerp system}

\author[Sartori et al.]{Lia F. Sartori$^{1}$\thanks{E-mail: lia.sartori@phys.ethz.ch}, Kevin Schawinski$^{1}$, Michael J. Koss$^{1,2}$, Claudio Ricci$^{3,4,5}$, 
\newauthor Ezequiel Treister$^{3}$, Daniel Stern$^{6}$, George Lansbury$^{7}$, W. Peter Maksym$^{8}$,
\newauthor Mislav Balokovi\'{c}$^{8,9}$, Poshak Gandhi$^{10}$, William C. Keel$^{11}$, David R. Ballantyne$^{12}$
\\
$^{1}$Institute for Particle Physics and Astrophysics, ETH Z\"{u}rich, Wolfgang-Pauli-Strasse 27, CH-8093 Z\"{u}rich, Switzerland,\\
$^{2}$Eureka Scientific Inc., 2452 Delmer St. Suite 100, Oakland, CA 94602, USA, \\
$^{3}$Instituto de Astrof\'{i}sica, Facultad de F\'{i}sica, Pontificia Universidad Cat\'{o}lica de Chile, Casilla 306, Santiago 22, Chile, \\
$^{4}$Kavli Institute for Astronomy and Astrophysics, Peking University, Beijing 100871, China, \\
$^{5}$Chinese Academy of Sciences South America Center for Astronomy and China-Chile Joint Center for Astronomy, \\
Camino El Observatorio 1515, Las Condes, Santiago, Chile, \\
$^{6}$Jet Propulsion Laboratory, California Institute of Technology, 4800 Oak Grove Drive, Pasadena, CA 91109, USA, \\
$^{7}$Institute of Astronomy, University of Cambridge, Madingley Road, Cambridge, CB3 0HA, UK, \\
$^{8}$Harvard-Smithsonian Center for Astrophysics, 60 Garden St., Cambridge, MA 02138, USA, \\
$^{9}$Cahill Center for Astronomy and Astrophysics, California Institute of Technology, Pasadena, CA 91125, USA, \\
$^{10}$Department of Physics and Astronomy, University of Southampton, Highfield, Southampton SO17 1BJ, UK, \\
$^{11}$Department of Physics and Astronomy, University of Alabama, Box 870324, Tuscaloosa, AL 35487, USA, \\
$^{12}$Center for Relativistic Astrophysics, School of Physics, Georgia Institute of Technology, Atlanta, GA 30332, USA
\\
}

\date{Accepted XXX. Received YYY; in original form ZZZ}

\pubyear{2016}

\begin{document}
\label{firstpage}
\pagerange{\pageref{firstpage}--\pageref{lastpage}}
\maketitle

\begin{abstract}

We present new {\it{Nuclear Spectroscopic Telescope Array}} ({\it{NuSTAR}}) observations of the core of IC\,2497, the galaxy associated with {\it{Hanny's Voorwerp}}. The combined fits of the \chandra (0.5-8 keV) and \nustar (3-24 keV) X-ray spectra, together with \wise mid-IR photometry, optical longslit spectroscopy and optical narrow-band imaging, suggest that the galaxy hosts a Compton-thick AGN ($N_{\rm H} \sim 2 \times 10^{24}$\,cm$^{-2}$, current intrinsic luminosity $L_{\rm bol} \sim 2-5 \times 10^{44}$\,erg\,s$^{-1}$) whose luminosity dropped by a factor of $\sim$\textcolor{black}{\rm{50}} within the last $\sim 100$ kyr. This corresponds to a change in Eddington ratio from $\rm \lambda_{Edd} \sim$ \textcolor{black}{\rm{0.35}} to $\rm \lambda_{Edd} \sim$ \textcolor{black}{\rm{0.007}}. We argue that the AGN in IC\,2497 should not be classified as a changing-look AGN, but rather we favour the interpretation where the AGN is undergoing a change in accretion state (from radiatively efficient to radiatively inefficient). In this scenario the observed drop in luminosity and Eddington ratio corresponds to the final stage of an AGN accretion phase. Our results are consistent with previous studies in the optical, X-ray and radio although the magnitude of the drop is lower than previously suggested. In addition, we discuss a possible analogy between X-ray binaries and an AGN.

\end{abstract}




\section{Introduction}

The discovery of {\it{Hanny's Voorwerp}} (HV hereafter) by a citizen scientist taking part in the Galaxy Zoo project (\citealt{Lintott2008,Lintott2009}) opened the field for studying Active Galactic Nuclei (AGN) variability on $10^{4}-10^{5}$ yr time-scales.
HV is an extended emission line region ($11 \times 16$ kpc in projected extent) located at $\sim 20$ kpc in projection from the core of IC\,2497, a massive ($M_* \sim 10^{11} M_{\odot}$), nearby ($z = 0.0502$) postmerger galaxy. The optical spectrum and the Sloan Digital Sky Survey (SDSS) $g$-band luminosity of HV are dominated by [$\mathrm{O}$\textsc{iii}]$\lambda 5007$ emission. The presence of emission lines of high-ionization species such as [$\mathrm{Ne}$\textsc{v}]$\lambda \lambda 3346, 3426$ and [\heii]$\lambda4616$, together with the relatively quiet kinematics (line widths $<100$ km s$^{-1}$) and the low electron temperature ($T_{e} \sim 13,500$K) suggests that the gas is photoionized by the hard continuum of an AGN in IC\,2497 rather than ionization from star formation or shocks \citep{Lintott2009}.
In order to produce sufficient ionizing photons to power the observed emission, the AGN output should be at least $L_{\rm ion} \sim 10^{45}$\,erg\,s$^{-1}$ between 1 and 4 Ryd (13.6-54.4 eV), corresponding to a bolometric luminosity $L_{\rm bol} \sim 10^{46}$\,erg\,s$^{-1}$.

Although the photoionization state of HV requires the presence of a strong AGN, the optical spectrum of the core of IC\,2497 only shows weak optical emission lines (\citealt{Lintott2009}; \citealt{Keel2012b}) and a weak radio source ($\sim10^{38}$\,erg\,s$^{-1}$; \citealt{Jozsa2009}). The two possible explanations to reconcile the observed and expected AGN output are: (1) the dust in the central part of the galaxy is arranged in such a way that the quasar appears obscured only along our line of sight but not in the direction of HV, or (2) the AGN dropped in luminosity in the last $\sim 100$ kyr, the travel time needed for the ionizing photons to reach the cloud \textcolor{black}{{\rm{(we note that this time is a lower limit since the measured distance is a projected distance)}}}. The first scenario was favoured by \cite{Jozsa2009}: using multifrequency radio observations they found that HV is embedded in a massive $\sim 10^9 M_{\odot}$ $\mathrm{H}$\textsc{i} reservoir crossed by an outflow connecting the centre of the galaxy to HV (see also \citealt{Rampadarath2010}), and derived a lower limit for the nuclear $\mathrm{H}$\textsc{i} column density of $2.8 \pm 0.4 \times 10^{21}$ cm$^{-2}$. Based on these data they suggested a scenario where the AGN is strongly obscured and the outflow cleared a path for the ionizing radiation to escape in the direction of HV.
On the other hand, the fading scenario was supported by \cite{Lintott2009}, \cite{Schawinski2010b}, \cite{Keel2012b}, \cite{Sartori2016} and \cite{Keel2017}. First, the IR data from {\it{IRAS}} show no evidence of the reprocessed luminosity expected from an obscured quasar of the required luminosity. Secondly, the available soft X-ray \chandra and \xmm spectra are well fitted by an unabsorbed power law plus a component for the hot gas in the galaxy, and the quality of the fit is not improved by the addition of an obscured AGN component. Finally, {\it{Hubble Space Telescope}} ({\it{HST}}) imaging shows a complex dust structure in the centre of IC\,2497, but the view of the nucleus is not hindered by the detected dust lanes.

Despite intensive efforts to measure it, the intrinsic current luminosity of the AGN in IC\,2497 is still not known. In this paper we present new hard X-ray observations obtained with the {\it{Nuclear Spectroscopic Telescope Array}} ({\it{NuSTAR}}; \citealt{Harrison2013}). Due to the high sensitivity and the hard X-ray coverage of \nustar ($3-79$ keV) we are able, for the first time, to put strong constraints on the current intrinsic AGN luminosity and to test directly the fading and obscuration scenarios. \textcolor{black}{{\rm{The paper is organized as follows. In section 2 we describe the data and data reduction. In section 3 we describe the performed analysis and how we estimate the past and current AGN bolometric luminosity. The results are discussed in Section 4.}}}


\section{Observations}

\begin{table*}
 \begin{tabular}{lcccc}
  \hline
  \hline
  \noalign{\smallskip}
   Obs. \# & \textcolor{black}{\rm{Observatory}} & Observation date & Observation ID & Net exposure \\
   \noalign{\smallskip}
  \hline
  \noalign{\smallskip}
  1 & \chandra & 2012 Jan 8 & 13966 & 59.35 ks \\
  \noalign{\smallskip}
  2 & \chandra & 2012 Jan 11 & 14381 & 53.4 ks \\
  \noalign{\smallskip}
  3 & \nustar & 2016 Sep 29 & 60201017002 & 31.4 ks \\
  \noalign{\smallskip}
  \hline
 \end{tabular}
 \caption{X-ray observations log for the data analysed in this paper. The archival \chandra observations cover the soft part of the X-ray spectrum ($0.5-8.0$ keV) while the newly obtained \nustar observations probe the hard part of the spectrum ($3-24$ keV).}
 \label{tab:log}
\end{table*}

We present the analysis of new \nustar observations of IC\,2497 ($3-24$ keV, \textcolor{black}{\rm{see Section \ref{sec:NuSTAR}}}), together with previously obtained \chandra X-ray observations of the galaxy ($0.5-8.0$ keV), and optical longslit spectroscopy \textcolor{black}{{\rm{and narrow-band imaging}}} of both the galaxy and HV. Details about the \chandra and \nustar observations are listed in Table\,\ref{tab:log}.

\subsection{Chandra}
\chandra \citep{Weisskopf2000} observations of IC\,2497 were obtained with the Advanced CCD Imaging Spectrometer S-array (ACIS-S; \citealt{Garmire2003}) in very faint (VFAINT) time-exposure (TE) mode on 2012 January 8 (ObsID 13966, 60 ks) and 2012 January 11 (ObsID 14381, 55 ks; PI K. Schawinski, Cycle 13). We performed standard data reduction starting from the level 1 event files using the \ciao 4.7 software \citep{Fruscione2006} provided by the \chandra X-ray Center (CXC). For both exposures we extracted the spectra and generated the response curves from a circular aperture with a 4 arcsec radius around the centroid of the emission using the \ciao task \textsc{specextract}. For the background we considered three source-free apertures $\sim 20$\,arcsec away from the target. We grouped the spectra with a minimum of 3 photons per bin using the \heasoft task \textsc{grppha}. Details about the \chandra data, data reduction and spectral extraction are provided in \cite{Sartori2016}.

\subsection{NuSTAR} \label{sec:NuSTAR}
\nustar observations of IC\,2497 were obtained on 2016 September 29 (ID 60201017002, 39 ks; program 02041, PI L. Sartori, GO cycle-2). We processed the raw event files using the \nustar  Data Analysis Software package (\nustardas version 1.6.0)\footnote{\url{http://heasarc.gsfc.nasa.gov/docs/nustar/analysis/nustar_swguide.pdf}} and the calibration files from \textsc{CALDB} version 20160922. Because of the slightly elevated background event rates around the SAA we processed the data with the parameters \textsc{SAAMODE=strict} and  \textsc{TENTACLE=yes} as recommended by the \nustar Science Operations Center (SOC). We extracted the source and background spectra using the \textsc{nuproducts} task included in the \nustardas package with the appropriate response and ancillary files. For each focal plane module we extracted the source spectra using a circular aperture with 45 arcsec radius. We checked that no additional bright sources are detected by \chandra within this region. For the background we considered \textcolor{black}{{\rm{two source-free apertures with 60 arcsec radius on the same detector chip as the source}}}. Similar to the \chandra spectra, we grouped the \nustar spectra with a minimum of 3 photons per bin using the \heasoft task \textsc{grppha}. The obtained net counts and background counts are listed in Table \ref{tab:nu}. \textcolor{black}{\rm{For the analysis we limited the \nustar spectrum to 3-24 keV instead of 3-79 keV because of the drop in sensitivity and the heightened background above 24 keV (e.g. \citealt{Wik2014}).}}


\begin{table*}
 \begin{tabular}{lccc}
  \hline
  \hline
  \noalign{\smallskip}
    & FPMA & FPMB & FPMA + FPMB \\
   \noalign{\smallskip}
  \hline
  \noalign{\smallskip}
  Source net counts & $59.9^{+12.4}_{-11.4}$ & $84.6^{+13.2}_{-12.2}$ & $144.5^{+17.7}_{-16.7}$ \\
  \noalign{\smallskip}
  Background counts & $63.1^{+16.6}_{-15.9}$ & $58.4^{+17.8}_{-17.1}$ & $121.5^{+24.1}_{-23.3}$  \\
  \noalign{\smallskip}
  \hline
 \end{tabular}
 \caption{Source net counts and background counts in the 8-24 keV band, for each \nustar focal plane.}
 \label{tab:nu}
\end{table*}

\subsection{Optical data}


Optical longslit spectra passing through both the nucleus of IC\,2497 and HV were obtained with the double-spectrograph system at the 4.2m William Herschel Telescope (WHT) on La Palma and with the 3m Shane telescope at the Lick Observatory. All the spectra were obtained with a 2\,arcsec slit width and nearly identical position on the sky. Details about observations and data reduction are given in \cite{Lintott2009}.

\textcolor{black}{{\rm{Narrow-band H$\alpha$ images are obtained with the {\it{HST}} using the tunable ramp filters on the Advanced Camera for Surveys (ACS). Details about the data  and data processing are described in \cite{Keel2012b} and \cite{Keel2017}.}}}

\section{Analysis}

The goal of this analysis is to measure the level of obscuration and the current intrinsic bolometric luminosity of the AGN in IC\,2497. A comparison of the obtained current luminosity with the past luminosity inferred from the photoionization state of HV will allow us to probe if the quasar really dropped in luminosity within the last $\sim 100$ kyr as suggested by previous studies, and the magnitude of the drop.

\subsection{X-ray emission}
We fit the obtained spectra using {\textsc{Xspec}} \citep{Arnaud1996} with Cash statistics \citep{Gehrels1986}, and estimate best-fitting parameters and errors using the MCMC routine implemented in {\textsc{Xspec}}. Because of the low signal to noise of the \nustar observations, we first examine the \chandra observations alone. We then perform combined fits considering different models. We verified that no significant variability is present between the observations but allow a $10\%$ offset to account for different responses of the detectors and uncertainties in the cross-calibrations. Errors for the best-fitting parameters are quoted at the $90\%$ confidence level.

\begin{figure*}
\includegraphics[scale=0.5]{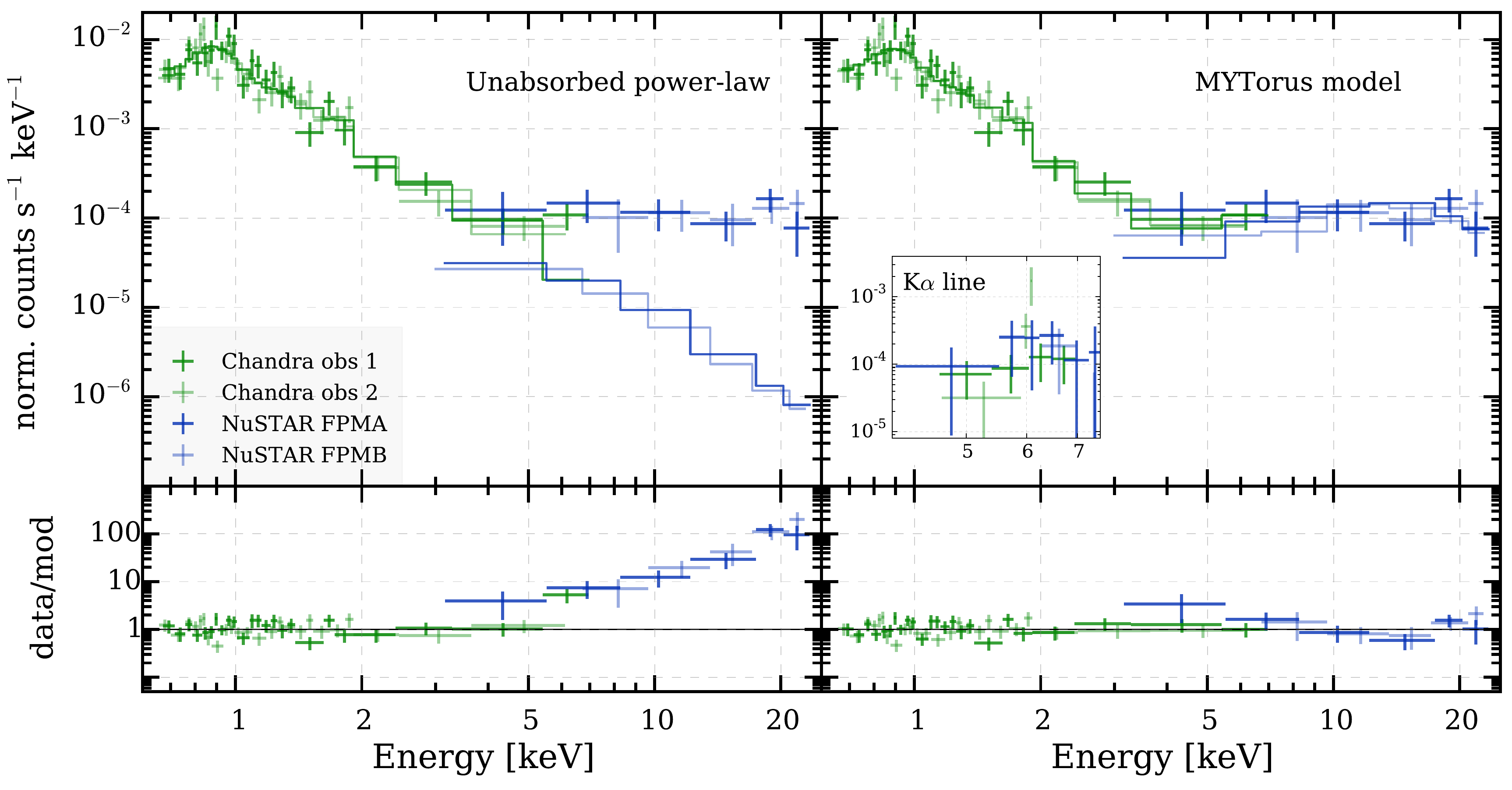}
\caption{Broadband X-ray spectrum from \chandra (green) and \nustar (blue). The crosses correspond to the data while the lines show the models. For plotting clarity we rebinned the spectrum with a minimum of 9 counts per bin. {\it{Left}}: The fit corresponds to the best fit to the \chandra data only, an unabsorbed power law with an APEC component, with extrapolation at higher energies. This model clearly underpredicts the hard X-ray counts observed with {\it{NuSTAR}}. {\it{Right}}: The fit corresponds to the \textsc{MYTorus} model with $\theta_{\rm i} = 90^{\rm o}$. This model is a better match to the data, and indicates that the source is \textcolor{black}{\rm{obscured. The box shows a zoom-in on the iron \Ka lline region with a different binning (minimum of 3 counts per bin)}}.}
\label{fig:spec}
\end{figure*}

\subsubsection{\chandra observations}

The $0.5-8.0$\,keV \chandra spectrum is best fitted (\textcolor{black}{\rm{C-stat/dof = 181.9/132}}) by an unabsorbed power law ($\Gamma = 2.14^{+0.27}_{-0.28}$), as expected for an unobscured AGN or for the Thomson-scattered emission of an obscured AGN, plus an APEC component representing the hot gas in the galaxy ($kT = 0.93^{+0.07}_{-0.11}$ keV). \textcolor{black}{\rm{We notice that the power law component may also be partially attributed to emission from star formation and X-ray binaries, as discussed in Section \ref{sec:SF}}}. The total absorption component is fixed to the Galactic value ($N_{\rm H} = 1.31 \times 10^{20}$\,cm$^{-2}$, as given by the Colden Calculator with the NRAO data set\footnote{\url{http://cxc.harvard.edu/toolkit/colden.jsp}}). A similar value for the photon index ($\Gamma = 2.23^{+0.57}_{-0.38}$) is obtained also when  the counts below 1.5 keV in order to avoid contamination by the hot gas. By fitting the $3.0-8.0$\,keV spectrum with a simple power law we obtain a photon index $\Gamma = -0.76^{+1.45}_{-0.05}$. This low value is an indication that the source is heavily obscured and that the power law observed in the soft X-ray spectrum corresponds to the scattered component. However, due to the narrow energy interval considered, adding an obscured component to our fiducial model for the \chandra data does not improve the quality of the fit significantly, so that a combined fit with \nustar is needed to determine the real amount of obscuration.

\subsubsection{Possible contribution from star formation}\label{sec:SF}

\textcolor{black}{\rm{IC 2497 is a luminous infrared galaxy (LIRG) with far-IR luminosity $L_{\rm FIR} = 6.6 \times 10^{44}$\,erg\,s$^{-1}$}. Assuming the FIR - star formation rate relation of \cite{Kennicutt1998}, this corresponds to a star formation rate $SFR \sim 30 M_{\odot}$\,yr$^{-1}$. Using the relations of \cite{Ranalli2003} or \cite{Lehmer2016}, the expected $2-10$ keV luminosity is $L_{\rm 2-10 keV} \sim 1.5 \times 10^{41}$ erg s$^{-1}$, which is comparable (within the large scatter of the correlations) to the observed $2-10$ keV luminosity measured from the \chandra and \nustar spectra (we stress that all the used relations have a large scatter, so that the results have to be considered as an order of magnitude estimation). In addition, the low {\it{W1}}-{\it{W2}} colour, {\it{W1}}-{\it{W2}} = 0.3, suggests that the mid-IR emission is dominated by stellar emission (see Section \ref{sec:mir} for a discussion of the mid-IR emission). Altogether, this points to a scenario where star formation provides an important contribution to the soft part of the X-ray spectrum. We will consider this component in the analysis below.  }

\subsubsection{Phenomenological model} \label{sec:PE}

Applying the unabsorbed power law model described above to the $0.5-24.0$\,keV spectrum significantly underpredicts the hard X-ray counts observed with {\it{NuSTAR}}, especially in the $10.0-20.0$\,keV range \textcolor{black}{\rm{(see Fig. \ref{fig:spec}, left)}}. This again is an indication that the source is highly obscured (e.g. \citealt{Koss2016}; \citealt{Ricci2016}). To infer the intrinsic column density $N_{\rm H}$ and the X-ray luminosity of the source we first fit the $0.5-24.0$\,keV spectrum with a phenomenological model which includes 1) an absorbed power law representing the intrinsic emission of the AGN, where we consider both photoelectric absorption and Compton scattering, 2) \textcolor{black}{\rm{a second unabsorbed power law representing the scattered AGN component and the contribution from star formation}}, 3) the unabsorbed reflected X-ray emission from neutral material above the accretion disc (the PEXRAV model; \citealt{Magdziarz1995}), 4) a Gaussian representing the narrow \Ka lline usually detected in AGN (e.g. \citealt{Nandra1994}; \citealt{Shu2010}; \citealt{Ricci2014}), with fixed energy $E = 6.4$keV and sigma $\sigma = 10$eV, and 5) an APEC model for the hot gas in the galaxy: \\

\textsc{phabs$_{\rm GAL}$} $\times$ (\textsc{zphabs} $\times$ \textsc{cabs} $\times$ \textsc{zpow} $+$ \textsc{pexrav} $+$ \textsc{zpow} $+$ \textsc{zgauss} $+$ \textsc{apec}). \\


\textcolor{black}{\rm{The photon index of the intrinsic power law is fixed at $\Gamma = 1.9$ as expected for an average AGN (e.g. \citealt{Nandra1994}; \citealt{Piconcelli2005}; \citealt{Ricci2015}), while the photon index of the unabsorbed power law is let free to vary. The best fit results in a column density $N_{\rm H} = 1.87^{+1.27}_{-0.21} \times 10^{24}$ cm$^{-2}$ and an intrinsic 2-10 keV luminosity (intended as the luminosity of the intrinsic power law) $L_{\rm 2-10 keV} = 2.3 \times 10^{43}$ erg s$^{-1}$. We note that the reflection parameter $R$ is an upper limit and the best-fitting parameters do not change significantly if we fix $R = -1 \times 10^{-10}$. This suggests that the reflection component is not statistically significant.}}

In order to estimate the equivalent width (EW) of the narrow \Ka lline we fit the \chandra and \nustar data in the $3.0-24.0$\,keV range simultaneously with a simplified model: \textsc{zpow} $+$ \textsc{zgauss}. We obtained EW $ 1.74^{+0.86}_{-0.72}$ keV, which is again consistent with heavy obscuration (e.g. \citealt{Matt1996}).

\subsubsection{BNtorus model} \label{sec:BN}

A more physically motivated model for highly obscured sources is the \textsc{BNtorus} model developed by \cite{Brightman2011}, which considers both absorbed and reprocessed emission of an intrinsic power law from a torus. In order to fit the whole $0.5-24.0$\,keV spectrum simultaneously we add a scattered power law component with same photon index as the intrinsic one and an APEC model in a similar way as described above: \\

\textsc{phabs$_{\rm GAL}$} $\times$ (\textcolor{black}{\textsc{BNtorus}} $+$ \textsc{zpow} $+$ \textsc{apec}).\\

In order to obtain a good fit we need to fix the inclination angle of the torus to the maximum permitted value $\theta_{\rm i} = 87.1^{\rm o}$. Varying the opening angle of the torus between $\theta_{\rm oa} = 40^{\rm o}$ and $\theta_{\rm oa} = 80^{\rm o}$ provides fit parameters which are consistent within the uncertainties and statistically indistinguishable. For $\theta_{\rm oa} = 60^{\rm o}$ the fit implies \textcolor{black}{\rm{$N_{\rm H} = 2.10^{+1.05}_{-0.57} \times 10^{24}$ $\rm {cm}^{-2}$}} and \textcolor{black}{\rm{$L_{\rm 2-10 keV} = 1.0 \times 10^{43}$ erg $\rm s^{-1}$}}.

\subsubsection{MYtorus model} \label{sec:MY}

A second physically motivated model for highly obscured AGN is the \textsc{MYtorus} model developed by \cite{Murphy2009}, which includes three components: the intrinsic absorbed power law continuum (\textsc{MYtorusZ}), the scattered continuum (\textsc{MYtorusS}) and the line emission (\textsc{MYtorusL}). In order to avoid artefacts present in the \textsc{MYtorus} model at low energies we cut the spectrum at 0.65 keV instead of 0.5 keV. As before, we add a scattered power law and an APEC component so that the total model is: \\

\textsc{phabs$_{\rm GAL}$} $\times$ (\textsc{MYtorusZ} $\times$ \textsc{zpow} $+$ \textsc{MYtorusS} $+$ \textsc{MYtorusL} $+$ \textsc{zpow} $+$ \textsc{apec}). \\

\begin{figure}
\includegraphics[scale=0.44]{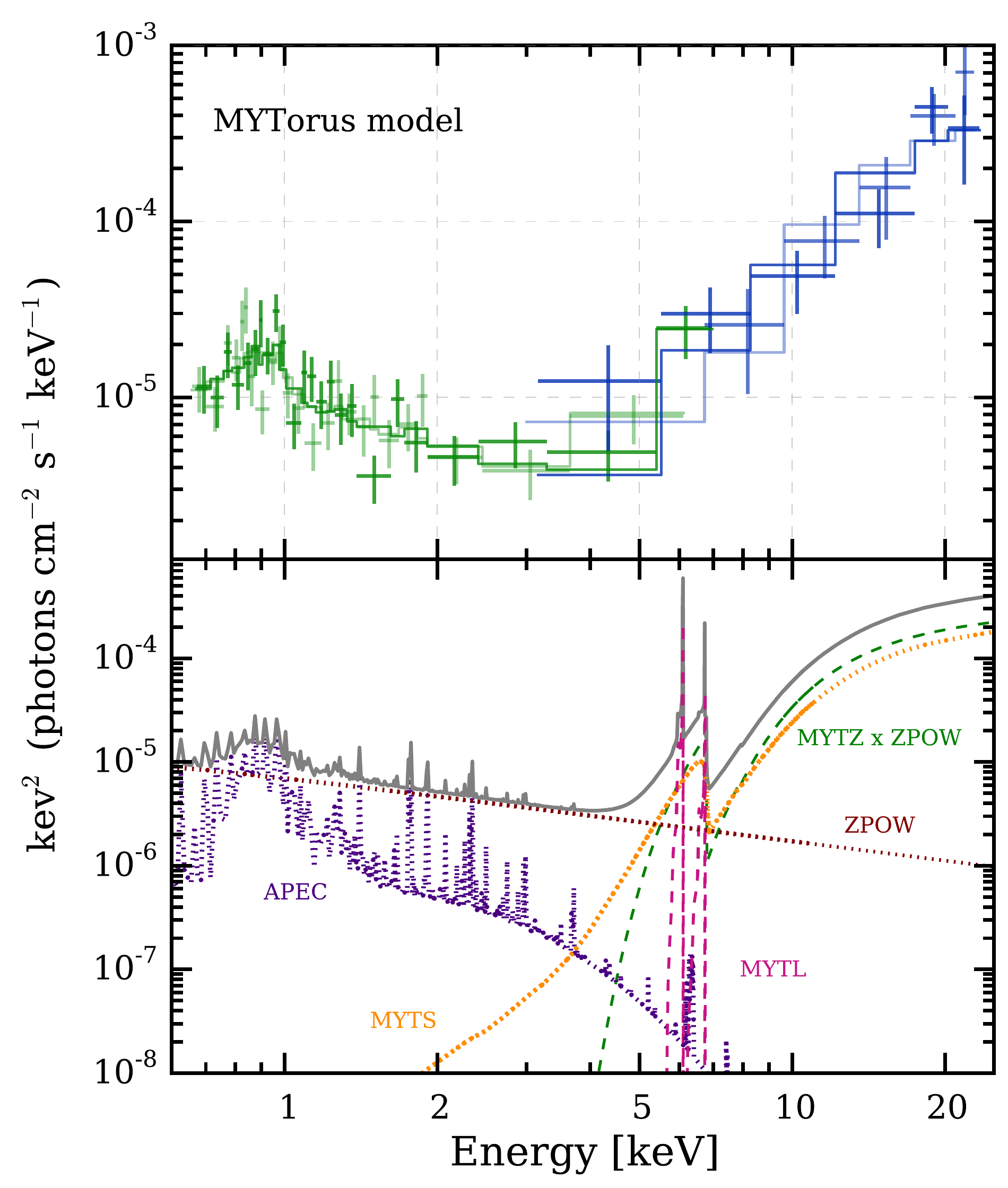}
\caption{\textcolor{black}{\rm{{\it{Top}}: same as Fig. \ref{fig:spec}, right, but for the unfolded spectrum. {\it{Bottom}}: unfolded model components for the best-fitting \textsc{MYTorus} model with $\theta_{\rm i} = 90^{\rm o}$.}}}
\label{fig:spec_2}
\end{figure}

\textcolor{black}{\rm{As in the previous model, the photon index of the intrinsic power law is fixed at $\Gamma = 1.9$ while the photon index of the unabsorbed power law is let free to vary. The power law parameters of the MYTorus components are tied together.}} Due to the low signal to noise of the spectrum, the column density and normalizations of the different \textsc{MYtorus} components are considered to be the same \textcolor{black}{\rm{(the available data do not allow us to constrain a different normalization of the scattered and line components with respect to the zeroth-order continuum represented by \textsc{MYtorusZ})}}. In addition, we fix the inclination angle to $\theta_{\rm i} = 90^{\rm o}$ \textcolor{black}{\rm{(Fig. \ref{fig:spec}, right, and Fig. \ref{fig:spec_2})}} and $\theta_{\rm i} = 75^{\rm o}$. The fits return \textcolor{black}{\rm{$N_{\rm H} \sim 2 \times 10^{24}$ cm$^{-2}$}} and \textcolor{black}{\rm{$L_{\rm 2-10 keV} \sim 1.2 \times 10^{43}$ erg $\rm s^{-1}$}} (see Table \ref{tab:res} for the values corresponding to the individual fits).

\begin{table*}
 \begin{tabular}{lcc}
  \hline
  \hline
  \noalign{\smallskip}
  \multicolumn{3}{c}{Phenomenological model (Sec. \ref{sec:PE})}\\
  \noalign{\smallskip}
  \hline
  \noalign{\smallskip}
  Column density $N_{\rm H}$ ($10^{24}$ $\rm cm^{-2}$) & \multicolumn{2}{c}{$1.87^{+1.27}_{-0.21}$}\\
  \noalign{\smallskip}
  Reflection parameter $R$ & \multicolumn{2}{c}{$\le 0.02$}\\
  \noalign{\smallskip}
  Plasma temperature $kT$ (keV) & \multicolumn{2}{c}{$0.95^{+0.1}_{-0.08}$}\\
  \noalign{\smallskip}
  Soft $\Gamma$ & \multicolumn{2}{c}{$2.32^{+0.5}_{-0.14}$}\\
  \noalign{\smallskip}
  C-stat/dof & \multicolumn{2}{c}{$246.8/222$}\\
  \noalign{\smallskip}
  Intrinsic 2-10 keV luminosity $L_{\rm 2-10 keV}$ (erg $\rm s^{-1}$) & \multicolumn{2}{c}{$(2.3^{+4.4}_{-0.8}) \times 10^{43}$}\\
  \noalign{\smallskip}
  Bolometric luminosity $L_{\rm bol}$ (erg $\rm s^{-1}$) & \multicolumn{2}{c}{$(4.9^{+14.9}_{-2.1}) \times 10^{44}$}\\
  \noalign{\smallskip}
  \hline
  \hline
  \noalign{\smallskip}
  \multicolumn{3}{c}{BNtorus model (Sec. \ref{sec:BN})}\\
  \noalign{\smallskip}
  \hline
  \noalign{\smallskip}
  Column density $N_{\rm H}$ ($10^{24}$ $\rm cm^{-2}$) & \multicolumn{2}{c}{$2.10^{+1.05}_{-0.57}$}\\
  \noalign{\smallskip}
  Inclination angle $\theta_{\rm i}$$^{\rm (1)}$ & \multicolumn{2}{c}{$87.1^{\rm o}$}\\
  \noalign{\smallskip}
  Torus opening angle $\theta_{\rm {oa}}$$^{\rm (1)}$ & \multicolumn{2}{c}{$60^{\rm o}$}\\
  \noalign{\smallskip}
  Plasma temperature $kT$ (keV) & \multicolumn{2}{c}{$1.00^{+0.15}_{-0.1}$}\\
  \noalign{\smallskip}
  Soft $\Gamma$ & \multicolumn{2}{c}{$2.86^{+0.12}_{-0.33}$}\\
  \noalign{\smallskip}
  C-stat/dof & \multicolumn{2}{c}{$245.3/217$}\\
  \noalign{\smallskip}
  Intrinsic 2-10 keV luminosity $L_{\rm 2-10 keV}$ (erg $\rm s^{-1}$) & \multicolumn{2}{c}{$(1.0^{+0.4}_{-0.4}) \times 10^{43}$}\\
  \noalign{\smallskip}
  Bolometric luminosity $L_{\rm bol}$ (erg $\rm s^{-1}$) & \multicolumn{2}{c}{$(1.9^{+1.0}_{-0.8}) \times 10^{44}$}\\
  \noalign{\smallskip}
  \hline
  \hline
  \noalign{\smallskip}
  \multicolumn{3}{c}{MYtorus model (Sec. \ref{sec:MY})}\\
  \noalign{\smallskip}
  \hline
  \noalign{\smallskip}
   & $\theta_{\rm i} = 90^{\rm o}$ & $\theta_{\rm i} = 75^{\rm o}$ \\
  \noalign{\smallskip}
  Column density $N_{\rm H}$ ($10^{24}$ $\rm cm^{-2}$) & $1.81^{+0.70}_{-0.36}$ & $2.09^{+1.03}_{-0.49}$ \\
  \noalign{\smallskip}
  Plasma temperature $kT$ (keV) & $0.98^{+0.23}_{-0.10}$ & $1.01^{+0.15}_{-0.14}$ \\
  \noalign{\smallskip}
  Soft $\Gamma$ & $2.62^{+0.34}_{-0.32}$ & $2.90^{+0.09}_{-0.33}$ \\
  \noalign{\smallskip}
  C-stat/dof & $239.8/213$ & $241.9/213$ \\
  \noalign{\smallskip}
  Intrinsic 2-10 keV luminosity $L_{\rm 2-10 keV}$ (erg $\rm s^{-1}$) & $(1.3^{+1.1}_{-0.5}) \times 10^{43}$ & $(1.2^{+0.6}_{-0.4}) \times 10^{43}$ \\
  \noalign{\smallskip}
  Bolometric luminosity $L_{\rm bol}$ (erg $\rm s^{-1}$) & $(2.4^{+2.8}_{-1.0}) \times 10^{44}$ & $(2.1^{+1.4}_{-0.8}) \times 10^{44}$ \\
  \noalign{\smallskip}
  \hline
 \end{tabular}
 \caption{Summary of the joint \chandra and \nustar X-ray spectral analysis. Details about the different models are given in the text. $^{\rm (1)}$ Frozen parameters. \textcolor{black}{\rm{The errors on the fit parameters correspond to the $90\%$ confidence range. The intrinsic 2-10 keV luminosities are intended as the luminosity of the intrinsic power law.}}}
 \label{tab:res}
\end{table*}


\subsection{Luminosity estimation}\label{sec:lum}

\begin{figure*}
\includegraphics[scale=0.6]{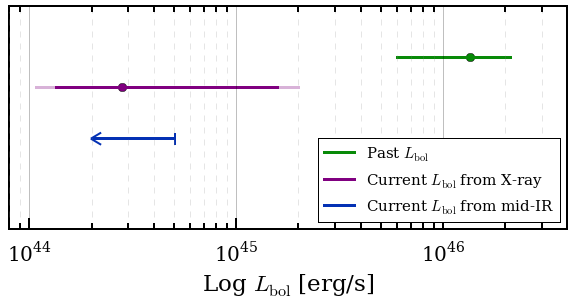}
\caption{\textcolor{black}{\rm{Past and current bolometric luminosity ranges as estimated from optical (green), X-ray (purple) and mid-IR (blue) data. The past luminosity range corresponds to the range between the minimum and maximum values obtained from optical data. The current bolometric luminosity range from X-ray data corresponds to the range obtained using the $68\%$ (strong purple) and $90\%$ (light purple) confidence intervals on the intrinsic power law normalization. The mid-IR data allows to compute an upper limit for the current bolometric luminosity. See Section \ref{sec:lum} for more details.}}}
\label{fig:lum}
\end{figure*}

\textcolor{black}{\rm{ In this section we estimate the current bolometric luminosity from the X-ray fits and from mid-IR data, and the past bolometric luminosity ($\sim 100$ kyr ago) using optical spectroscopy. The results are summarized in Fig. \ref{fig:lum}.}}

\subsubsection{Current bolometric luminosity from the X-ray spectrum}

The simultaneous \chandra and \nustar analysis of the  $0.5-24.0$\,keV spectrum of IC\,2497 suggests that the source is highly obscured, \textcolor{black}{\rm{$N_{\rm H} = 1.8-2.1 \times 10^{24}$ cm$^{-2}$}}, with intrinsic 2-10 keV luminosity \textcolor{black}{\rm{ (luminosity of the intrinsic power law) $L_{\rm 2-10 keV} = 1 - 2.3 \times 10^{43}$ erg s$^{-1}$}}. 

\textcolor{black}{\rm{Assuming the luminosity dependent X-ray to bolometric correction from \cite{Marconi2004}, the measured 2-10 keV luminosity corresponds to a bolometric luminosity $L_{\rm bol} = 1.9-4.9 \times 10^{44}$ erg $\rm s^{-1}$.}}

\textcolor{black}{\rm{Fig. \ref{fig:lum} shows the range covered by the estimated bolometric luminosity considering also the uncertainties on the intrinsic power law. For every model we compute the 2-10 keV luminosity for the best-fitting normalization and for the minimum and maximum value allowed from the $90\%$ and $68\%$ confidence regions. We then infer the bolometric luminosity as explained above. In the plot we show the mean value obtained from the best-fit parameters, and the minimum and maximum values for the two confidence levels measured with the different models.}}

\subsubsection{Current bolometric luminosity from mid-IR emission}\label{sec:mir}

The {\it{WISE}} observations \citep{Wright2010} of IC\,2497 allow us to compute an upper limit to the current bolometric luminosity of the obscured quasar. First, the {\it{W1}}-{\it{W2}} colour is low, {\it{W1}}-{\it{W2}} = 0.3\footnote{We checked that no variability is present between {\it{WISE}} and {\it{NEOWISE}} (\citealt{Mainzer2011}).}, which suggests that the mid-IR flux is dominated by stellar emission (e.g. \citealt{Stern2005}; \citealt{Jarrett2011}; \citealt{Stern2012}). Moreover, the emission detected with {\it{W3}} and {\it{W4}} (12 and 22 $\mu$m, respectively) appears effectively not to be absorbed e.g. by dust also in the case of highly obscured quasars (e.g. \citealt{Gandhi2009}; \citealt{Asmus2015}), so that the observed luminosity corresponds to the intrinsic one. Assuming a radio quiet quasar SED template from \cite{Elvis1994} and normalising it to the lowest non absorbed {\it{WISE}} point, i.e. {\it{W3}}, we obtain an upper limit for the bolometric luminosity $L_{\rm bol} = 5.0 \times 10^{44}$ erg $\rm s^{-1}$. This value is consistent with that obtained from the X-ray analysis.

\subsubsection{Past bolometric luminosity from optical emission}

A detailed analysis of the optical longslit spectra, which allow the computation of the lower and upper limits to the AGN output needed to photoionize HV, is presented in \cite{Lintott2009}. Here we provide a brief summary of their findings. A lower limit for the required AGN output is obtained from simple recombination balance, i.e. by assuming that the number of ionizations is equal to the number of recombinations, and that the number of ionizing photons is high enough to power the observed recombination lines. The integrated $\rm H \beta$ luminosity of HV is $1.4 \times 10^{41}$ erg s$^{-1}$. Assuming typical nebular conditions and a flat ionizing continuum ($F_{\nu} \propto \nu^{-1}$) this corresponds to a lower limit for the ionizing luminosity $L_{\rm ion} > 1 \times 10^{45}$\,erg\,s$^{-1}$. On the other hand, the ionization parameter obtained from the [$\mathrm{O}$\textsc{ii}]$\lambda 3727$/[$\mathrm{O}$\textsc{iii}]$\lambda 5007$ ratio of HV is log(U)\,=\,$-2.2$ while the [$\mathrm{S}$\textsc{ii}]$\lambda 6717 /6731$ doublet ratio provides an upper limit for the electron density $n_{\rm e} < 50$\,cm$^{-3}$. This provides an upper limit for the ionizing photon density $n_{\rm phot} < 0.32$ cm$^{-3}$. Assuming a flat ionizing continuum and the mean projected separation between IC\,2497 and HV, $d=20$ kpc, the photon density obtained corresponds to an upper limit to the ionizing luminosity $L_{\rm ion} < 3.2 \times 10^{45}$\,erg\,s$^{-1}$.

\textcolor{black}{\rm{A similar analysis was performed also in \cite{Keel2017}. Based on {\it{HST}} narrow-band H$\alpha$ images and photoionization balance, they infer a lower limit (because of possible dust obscuration and/or a clumpy gas structure) for the isotropic emission rate of ionizing photons $Q_{\rm ion} \sim 3 \times 10^{55}$ photons s$^{-1}$ for almost the whole time range probed by the photoionized cloud, which again corresponds to a lower limit for the ionizing luminosity $L_{\rm ion} > 1.1 \times 10^{45}$\,erg\,s$^{-1}$.}}

\textcolor{black}{\rm{Altogether, the observations point to a past ionizing luminosity in the range $L_{\rm ion} = 1 - 3.2 \times 10^{45}$\,erg\,s$^{-1}$. By scaling a radio quasar template from \citealt{Elvis1994} to match the estimated ionizing luminosity between 1 and 4 Ryd, we infer a past bolometric luminosity $L_{\rm bol} = 0.6 - 2.1 \times 10^{46}$\,erg\,s$^{-1}$.}}

\section{Discussion}
We used new hard X-ray \nustar observations of IC\,2497, together with archival \chandra observations and {\it{WISE}} fluxes, to investigate the current state of its quasar. The data rule out the possibility that the quasar is currently radiating \textcolor{black}{at a rate} $L_{\rm bol} \sim 10^{46}$ erg s$^{-1}$ as required by the photoionization state of Hanny's Voorwerp, and suggest a drop in luminosity by a factor of $\sim$\textcolor{black}{\rm{50}} in the last $\sim$ 100 kyr. In addition, the newly obtained observations suggest that the source is Compton thick with $N_{\rm H} \sim 2 \times$\,{\textcolor{black}{$10^{24}$}}\,cm$^{-2}$. However, the obscuration is not the explanation for the discrepancy between the needed and observed levels of AGN emission; this suggests that significant AGN variability is required to explain the properties of this system. These findings also support the idea first proposed by \cite{Lintott2009} that Hanny's Voorwerp is a quasar ionization echo, although the magnitude of the drop in luminosity is significantly lower than previously thought (factor $\sim$ \textcolor{black}{\rm{50}} instead of $> 100$, \textcolor{black}{\rm{assuming the mean current and past luminosities obtained with our analysis}}).

Assuming a stellar mass = $M_* \sim 1.7 \times 10^{11} M_\odot$ as estimated from the K band magnitude and the $M_* - M_{BH}$ relation in \cite{Jahnke2009}, \textcolor{black}{\rm{as well as the mean current and past luminosities obtained with our analysis}}, the measured drop in bolometric luminosity corresponds to a drop in Eddington ratio (ER) from $\rm \lambda_{Edd} \sim$ \textcolor{black}{\rm{0.35}} to $\rm \lambda_{Edd} \sim$ \textcolor{black}{\rm{0.007}}. The ER found for luminous AGN is typically $\rm \lambda_{Edd} \sim 0.01 - 1$ (e.g. \citealt{Panessa2006}; \citealt{Kollmeier2006}). An ER $\rm \lambda_{Edd} \sim 0.007$ is more consistent with the predictions for a radiative inefficient (RIAF) or advection-dominated accretion flow (ADAF; e.g. \citealt{Narayan1995}; \citealt{Narayan1998}; \citealt{Ho2009}), where most of the energy is radiated in kinetic form. The drop in ER could therefore suggest that IC\,2497 is entering a regime where we might expect a change from a radiatively efficient to a radiatively inefficient accretion state (e.g. \citealt{Done2007}; \citealt{Alexander2012}). This hypothesis is supported also by the fact that IC\,2497 and its quasar show other characteristics common to ADAF sources, such as the presence of a compact radio source or jet, and the prevalence of low-ionization nebular conditions (e.g. \citealt{Terashima2002}; \citealt{Ptak2004}; \citealt{Ho2009}).

Accretion state changes are common in X-ray binaries. In these systems the switch from a radiatively efficient to a radiatively inefficient state is often linked to changes in the jet activity, and provides a source of mechanical feedback (e.g. \citealt{Done2007}; \citealt{Pakull2010}). As observed by \cite{Jozsa2009} and \cite{Rampadarath2010}, IC\,2497 shows a large scale radio jet which connects the galaxy to Hanny's Voorwerp, and a jet hotspot $\sim 300$ pc away from the AGN which may \textcolor{black}{\rm{correspond}} to a jet launched after the quasar started dropping in luminosity. A hypothesis is that the nuclear jet is connected to the current accretion state change, in analogy to what is observed in X-ray binaries, while the large scale radio jet was launched during a previous AGN phase. In fact, as suggested by \cite{Schawinski2015}, AGN may ``flicker" on and off 100-1000 times, with a typical AGN phase lasting $\sim 10^5$ yr, so that the host galaxy may present signatures related to different AGN phases. The idea of the AGN currently doing mechanical feedback on the galaxy is supported also by the presence of a nuclear outflow and a nuclear ring of expanding gas observed in the optical {\it{HST}} images (\citealt{Keel2012b}) and of a bubble expanding in the hot gas around the AGN seen in the \chandra data (\citealt{Sartori2016}). The analogy between IC\,2497 and what is observed in X-ray binaries provides further indication of a possible unification of black hole accretion physics from stellar mass BHs to SMBHs. However, it is important to notice that different time-scales have to be taken into account since the dynamical and viscous time-scales increase with black hole mass. If we assume the underlying physics for black hole accretion to be the same, the analogous of a state transition lasting $\sim 1$h for an $\sim 10M_{\odot}$ BH would last $\sim 10^{4-5}$yr for a $\sim 10^{9} M_{\odot}$ SMBH (see \citealt{Schawinski2010b}, \citealt{Sartori2016} and references therein for a more detailed description).

Other examples of AGN variability are given by the so-called {\it{changing-look AGN}}, which present a change in the AGN type (e.g. from Type 1 to Type 1.9) due to broadening or narrowing of the Balmer lines{\footnote{In this \textcolor{black}{\rm{section}} we refer to optical changing-look AGN, while other studies describe systems where the X-ray changing-look behaviour, from Compton thick to Compton thin or viceversa, is due to a change in the obscuration (e.g. \cite{Risaliti2005}; \citealt{Ricci2016b})} (\citealt{Denney2014}; \citealt{LaMassa2015}; \citealt{Ruan2015}; \citealt{MacLeold2016}; \citealt{Runnoe2016}; \citealt{Husemann2016}; \citealt{McElroy2016}; \citealt{Gezari2016}; \textcolor{blue}{Stern et al. submitted}). The appearance or disappearance of broad emission lines is often accompanied by a change in luminosity of a factor $\sim$10 over $\sim$10 yr time-scales. As described above, these time-scales are much shorter compared to time-scales expected for accretion state changes (e.g. \citealt{Sobolewska2011}; \citealt{Hickox2014}), and possible alternative explanations of the changing-look behaviour are variable absorption due to a clumpy torus (e.g. \citealt{Elitzur2012}), transient events, e.g. tidal disruption of a star by the central black hole (e.g. \citealt{Eracloeous1995}; \citealt{Merloni2015}), or major changes in the photoionization balance. The magnitude of the drop in luminosity measured in IC\,2497 is \textcolor{black}{\rm{at least}} a factor $\sim 2$ higher than what has been observed in changing-look AGN. Moreover, the \chandra and \nustar data do not show significant variability, and the upper limits obtained from archival {\it{WISE}}, {\it{NEOWISE}} and {\it{IRAS}} data seem to exclude that the total drop in luminosity happened within the last decades. For these reasons, we argue that the AGN in IC\,2497 should not be classified as a changing-look AGN. On the other hand, we suggest that changing-look AGN correspond to a short-time ($\sim 10-100$ yr) variability which is superimposed on the long-term ($\sim 10^{5-6}$ yr) AGN phases suggested by this work and other observations (e.g. \citealt{Schawinski2015}), high resolution (sub-kpc) simulations (\citealt{Hopkins2010}; \citealt{Novak2011}; \citealt{Bournaud2011}; \citealt{Gabor2013}; \citealt{DeGraf2014}; \citealt{Sijacki2015}) and theoretical models (\citealt{Siemiginowska1997}; \citealt{Sanders1981}; \citealt{DiMatteo2005}; \citealt{Hopkins2005}; \citealt{Springel2005}; \citealt{King2007}; \citealt{King2015}).

\section{Conclusions}
In this paper we present new \nustar observations of IC\,2497, the galaxy associated with Hanny's Voorwerp. The analysis of these new data, together with archival \chandra X-ray observations, optical longslit spectroscopy \textcolor{black}{{\rm{and narrow-band imaging}}}, and \wise mid-IR photometry, shows that the AGN in IC\,2497 is Compton thick, and its luminosity dropped by a factor of $\sim$\textcolor{black}{\rm{50}} within the last $\sim 100$ kyr. These findings support the idea that Hanny's Voorwerp is an ionization echo, although the magnitude of the quasar's drop in luminosity is significantly lower than previously thought. On the other hand, the observed change in Eddington ratio, from $\rm \lambda_{Edd} \sim$ \textcolor{black}{\rm{0.35}} to $\rm \lambda_{Edd} \sim$ \textcolor{black}{\rm{0.007}}, suggests that the quasar in IC\,2497 is entering a regime where it is switching from a radiatively efficient to a radiatively inefficient accretion state, in a similar way as observed in X-ray binaries.

\section*{Acknowledgements}

\textcolor{black}{\rm{We thank the anonymous referee for helpful comments which improved the quality of the manuscript. We also thank the \nustar Cycle 2 TAC for the \nustar data on which this paper is based.}}
LFS and KS gratefully acknowledge support from Swiss National Science Foundation Grants PP00P2\_138979 and PP00P2\_166159. MJK acknowledges support from the Swiss National Science Foundation through the Ambizione fellowship grant PZ00P2\_154799/1. MJK acknowledges support from NASA through ADAP award NNH16CT03C. ET acknowledge support from: CONICYT-Chile grants Basal-CATA PFB-06/2007 and FONDECYT Regular 1160999. CR acknowledges the China-CONICYT fund, FONDECYT 1141218 and Basal-CATA PFB--06/2007. WPM acknowledges partial support by HST guest observer programme \#14271  under NASA contract NAS 5-26555 (HST) and Chandra Award GO5-16101X under NASA contract NAS8-03060 (CXC). GL acknowledges support from a Herchel Smith Research Fellowship of the University of Cambridge.  This work is based on observations with the \chandra satellite and has made use of software provided by the \chandra X-ray Center (CXC) in the application package \textsc{CIAO}. This work made use of data from the \nustar mission, a project led by the California Institute of Technology, managed by the Jet Propulsion Laboratory, and funded by the National Aeronautics and Space Administration. We thank the \nustar Operations, Software and Calibration teams for support with the execution and analysis of these observations. This research has made use of the \nustar Data Analysis Software (\textsc{NuSTARDAS}) jointly developed by the ASI Science DataCenter (ASDC, Italy) and the California Institute of Technology (USA). The research has made use of NASA's Astrophysics Data System Bibliographic Service. 










\bsp  
\label{lastpage}
\end{document}